\documentclass[aps,preprint]{revtex4}%
\usepackage{amsfonts}
\usepackage{amsmath}
\usepackage{amssymb}
\usepackage{graphicx}
\usepackage{pdfpages}
\usepackage{cleveref}
\usepackage{xcolor}
\usepackage{subcaption}%
\providecommand{\U}[1]{\protect\rule{.1in}{.1in}}

\begin{document}
\preprint{HEP/123-qed}
\title{Greybody factors of black holes in dRGT massive gravity coupled with nonlinear electrodynamics}
\author{Sara Kanzi}
\affiliation{}
\author{S. Habib Mazharimousavi}
\affiliation{}
\author{\.{I}zzet Sakall{\i}}
\affiliation{Physics Department, Arts and Sciences Faculty, Eastern Mediterranean
University, Famagusta, North Cyprus via Mersin 10, Turkey.}
\author{}
\affiliation{}

\begin{abstract}
In the context of the dRGT massive gravity coupled with nonlinear
electrodynamics, we present new dRGT black hole solutions. Together with the
thermodynamical properties of the solutions, we study the greybody factor of
the corresponding black hole solution. To this end, we compute the rigorous
bound on the greybody factor for the obtained dRGT black holes. The obtained
results are graphically represented for different values of the theory's
physical parameters. Our analysis shows that the charged dRGT black holes of
nonlinear electrodynamics evaporate quicker than the charged dRGT black
holes originated from linear electrodynamics.

\end{abstract}
\volumeyear{ }
\eid{ }
\date{\today}
\received{}

\maketitle
\tableofcontents

\section{Introduction}

For almost a century, the theory of general relativity (GR) has been known to describe the force of gravity in perfect harmony with observations. However, the unanswered questions like the interface between cosmology/gravity and particle physics such as the hierarchy problem, the cosmological constant problem, and the source of the late-time acceleration of Universe have boosted the researchers to seek for alternatives to GR theory. Moreover, it has been understood that GR cannot define Universe neither in ultraviolet nor in  infrared region. In other words, according to the observations of large scale phenomena, the gravity necessitates a vast effort and understanding.\par Today, one of the best modification theories is the massive gravity, which consists a massive graviton. Obviously, the perception of massive gravity theories is feasible with the help of gravitational wave (GW) astronomy \cite{T1}. In this regard, bounds on the massive gravity are obtained via gravitational wave emission \cite{T2,T3,T4} and other propagation/radiation mechanisms \cite{T5,T6}. In addition to this, the accurate observations of binary black holes (by the LIGO and Virgo detectors), can also constraint the massive gravity \cite{1,LIGOmass,LIGOmass2}.  Furthermore, it is believed that the massive graviton having the Hubble
scale may be responsible for the accelerated expansion of Universe \cite{2,3}.
An experimental detection of graviton is a three-pipe
problem, however theories of massive gravity have a number of pathologies
\cite{4,5}. In fact, massive gravity has a long and winding history dating back to the 1930s. First time, Fierz and Pauli (FP) developed the theory of a massive spin-$2$ field that propagates on a flat spacetime \cite{6,7}. FP massive gravity is also known as a unique linear theory without instabilities (ghost free) \cite{SARA2}. Basically, massive gravity is nothing but a theory that modifies general relativity by taking into account of some additional terms in the Einstein–Hilbert action. Massive gravity generates 5 degrees of freedom associated with the massive spin-2 graviton. Furthermore, massive gravity uses a non-vanishing graviton mass instead of the dark energy to modify gravity in the IR region  \cite{T7}. \par 

In the 1970s van Dam and Veltman and (independently) Zakharov found a peculiar feature of FP massive gravity: its anticipations do not uniformly fit to those of GR in the limit of vanishing mass \cite{9,10}. In particular, while at small scales (shorter than the Compton wavelength of the graviton mass), Newton's gravitational law is recovered, the bending of light is only three quarters of the result obtained in GR. This is known as the vDVZ discontinuity, which is cured classically by the nonlinear Vainshtein mechanism due to certain low scale strongly coupled interactions \cite{11}. But, Boulware and Deser (BD) \cite{8} showed that the
ghost reappears at the non-linear level. In 2010, a ghost-free non-linear extension of the FP action was proposed by de Rham-Gabadadze-Tolly (dRGT) 
 \cite{12,13,14}. 
 According to this theory, the sixth BD ghost mode is omitted by using a special type of potential to recover the Hamiltonian constraint \cite{15,16}, which is valid
in any dimensions. In this new massive gravity, adding mass to the graviton
does not change significantly the physics on a small scale from the GR, as it
was expected \cite{4}. There are numerous works published in dRGT theory which
lead a growing interest in this theory \cite{17}. For instance, some
cosmological and black hole solutions obtained in the dRGT theory\ can be seen
in Refs. \cite{42,43,44,45,46,47,48,49}
and\ \cite{50,51,52,53,54,55,56,57,58,59,60,61,62,63}, respectively. In \cite{T10} a general solution has been found for an isotropic reference metric.
In particular, the first non-trivial black hole solution of the $3+1$-dimensional
dRGT gravity with cosmological constant \cite{SARA3,65} was found by Vegh \cite{64}.\par 
\par Regarding to the definition of the Hawking radiation \cite{66} and greybody
factor, there are different methods to evaluate the transmission probability
and greybody factor, such as\ the WKB approximation, matching method
\cite{67,68,69,70,71,72,73}, and rigorous bound method \cite{74}. Studies
about greybody factors have been increasingly gaining attention in the
literature due to its observational evidence potential (see for example
\cite{18,19,20,21,22,23,24,25,26,27,28,29,30,31,32,33,34,35,36,37,38,39,40,41}
and references therein).\par In the present study, we first introduce the
$3+1$-dimensional black hole solutions in the dRGT massive gravity coupled
with nonlinear electrodynamics and then analyze their greybody factors with
the method of rigorous bound \cite{75}. As is well known, nonlinear electrodynamics is nothing but the extension of Maxwell's electromagnetic theory which comes out when the self-interactions in the field equations are allowed. In general, they can be constructed from the Lagrangian of a vector field that meets the following conditions: invariance under the Lorentz and $U(1)$ gauge groups and the Lagrangian should depend only on the combinations of the field and its first derivative \big( $\mathcal{L}=\mathcal{L}\left(A_{u}, \partial_{v} A_{u}\right)$\big). Historically, the nonlinear electrodynamics proposals date back to 1930s. In 1934, Born and Infeld proposed the Born–Infeld electrodynamics to dispose of the self-energy of a point charge \cite{rev1,rev2}. Although the Born-Infeld electrodynamics was initially seen as a fundamental theory for the electromagnetism, but it was later seen that  it was not renormalizable. For this reason, today it is only accepted as an effective theory. Afterwards, in 1936, Heisenberg and
Euler \cite{rev3} conjectured that the self-coupling of the electromagnetic field induced by virtual pairs of electrons-positrons having energies lower than the electron mass can be treated as a new effective field theory. Today, this theory is known as Euler-Heisenberg electrodynamics and it is the first nonlinear electromagnetic theory, which explains the vacuum polarization effect in the quantum electrodynamics (i.e., relativistic quantum field theory of electrodynamics) \cite{rev4}. \par Nonlinear electrodynamics have some remarkable properties comparing with the Maxwell's electrodynamics. Among them, its non-trivial effect on the \textit{radiation propagation} is very important. Namely, the electromagnetic field self-interacts generate deformations on  the light cone because of the nonlinear structure of the field equations \cite{rev5}. In other words, in the nonlinear electrodynamics context, the background field modifies the propagation speed of the electromagnetic waves and hence it gives rise to the birefringence phenomenon \cite{rev6}. Because of different motivations, various nonlinear electrodynamics theories (for instance logarithmic and exponential electrodynamics \cite{rev7,rev8,rev9}) were thrown out for consideration, and today the nonlinear electrodynamics has become a class
of electromagnetic theories \cite{rev10}. The nonlinear electrodynamics has many applications in several subjects including, for example, the black holes
\cite{rev11,rev12,rev13,rev14,rev15,rev16}, cosmology \cite{rev17,rev18}, optics \cite{rev19}, and even in biological systems \cite{rev20}. In this context, the main motivation of this study is to consider the  nonlinear electrodynamics in the dRGT theory and to examine its modification on the Hawking radiation. 

The remainder of this paper is organized as follows: Section II lays out the
$3+1$-dimensional black hole solutions in the dRGT massive gravity coupled
with nonlinear electrodynamics. In Sec. III, we consider the massless scalar
perturbations in the geometry of the $3+1$-dimensional black hole solutions in
the dRGT massive gravity coupled with nonlinear electrodynamics. Section
IV\ is devoted to the computation of the greybody factors with the method of
rigorous bound. Finally, we discuss our results and conclude in Sec. V. We
follow the metric signature $(-+++)$ and use the geometrized units, where
$G=c=1$.

\section{$3+1$-dimensional black hole solution in\ dRGT massive gravity
coupled with nonlinear electrodynamics}

In this section, we first consider the action of the dRGT massive gravity
without matter source and cosmological constant in $3+1-$dimensions, which is
given by \cite{12,13,14,SARA5,SARA1,SARA6}
\begin{equation}
S=\frac{1}{2}\int d^{4}x\sqrt{-g}\left[R\left(g\right) +{m}_{g}^{2}\left(\mathcal{U}_{2}+\mathcal{\alpha}_{3}\mathcal{U}_{3}+\mathcal{\alpha}_{4} \mathcal{U}_{4}\right)+\mathcal{L}  \right] ,  
\label{HM1}%
\end{equation}
in which
\begin{equation}
\mathcal{U}_{2}=Tr\left(  \mathcal{K}\right)  ^{2}-Tr\left(  \mathcal{K}%
^{2}\right)  , \label{H1}%
\end{equation}%
\begin{equation}
\mathcal{U}_{3}=Tr\left(  \mathcal{K}\right)  ^{3}-3Tr\left(  \mathcal{K}%
\right)  Tr\left(  \mathcal{K}^{2}\right)  +2Tr\left(  \mathcal{K}^{3}\right)
, \label{H2}%
\end{equation}
and%
\begin{equation}
\mathcal{U}_{4}=Tr\left(  \mathcal{K}\right)  ^{4}-6Tr\left(  \mathcal{K}%
^{2}\right)  Tr\left(  \mathcal{K}\right)  ^{2}+8Tr\left(  \mathcal{K}%
^{3}\right)  Tr\left(  \mathcal{K}\right)  +3Tr\left(  
19
 \mathcal{K}^{2}\right)
^{2}-6Tr\left(  \mathcal{K}^{4}\right)  . \label{H3}%
\end{equation}
Herein, $m_{g}$ is the mass of the graviton, $\alpha_{3}$ and $\alpha_{4}$ are
constants of the theory, and $\mathcal{K}$ represents a $4\times4$ matrix
defined by
\begin{equation}
\mathcal{K}_{\mu}^{\nu}=\mathcal{\delta}_{\mu}^{\nu}-\sqrt{g^{\alpha\gamma
}f_{\gamma\beta}}. \label{H4}%
\end{equation}
In latter equation, $g^{\alpha\gamma}$ is the inverse of the metric tensor and
$f_{\gamma\beta}$ is a symmetric tensor which is called reference (or
fiducial) metric. The nonlinear electrodynamics Lagrangian $\mathcal{L}$ is
defined by \cite{65}%
\begin{equation}
\mathcal{L}=\frac{-\mathcal{F}}{1-\frac{b}{\sqrt{8}}\sqrt{-\mathcal{F}}},
\label{H5}%
\end{equation}
where $b$ is a positive parameter and $\mathcal{F}=F_{\alpha\beta}%
F^{\alpha\beta}$ is nothing but the Maxwell invariant with a pure electric
field%
\begin{equation}
\mathbf{F}=E\left(  r\right)  dt\wedge dr. \label{H6}%
\end{equation}
Variation of the action with respect to electric potential admits the
following Maxwell nonlinear equation%
\begin{equation}
d\left(  \mathbf{\tilde{F}}\frac{d\mathcal{L}}{d\mathcal{F}}\right)  =0,
\label{H7}%
\end{equation}
where $\mathbf{\tilde{F}}$ is the dual of $\mathbf{F.}$ The variation of the
metric with respect to the metric tensor yields the following field equations%
\begin{equation}
G_{\mu}^{\nu}+m_{g}^{2}X_{\mu}^{\nu}=T_{\mu}^{\nu}, \label{H8}%
\end{equation}
in which
\begin{equation}
X_{\mu\nu}=\mathcal{K}_{\mu\nu}-\mathcal{K}g_{\mu\nu}-\alpha\left(
\mathcal{K}_{\mu\nu}^{2}-\mathcal{KK}_{\mu\nu}+\frac{\mathcal{U}_{2}}{2}%
g_{\mu\nu}\right)  +\label{H9}\\
3\beta\left(  \mathcal{K}_{\mu\nu}^{3}-\mathcal{KK}_{\mu\nu}^{2}%
+\frac{\mathcal{U}_{2}}{2}\mathcal{K}_{\mu\nu}-\frac{\mathcal{U}_{3}}{6}%
g_{\mu\nu}\right)  ,
\end{equation}
with%
\begin{equation}
\alpha=1+3\alpha_{3}, \label{H10}%
\end{equation}
and%
\begin{equation}
\beta=\alpha_{3}+4\alpha_{4}. \label{H11}%
\end{equation}
Furthermore, $T_{\mu}^{\nu}$ denotes the nonlinear electrodynamics energy
momentum tensor, which is given by%
\begin{equation}
T_{\mu}^{\nu}=\frac{1}{2}\left(  \mathcal{L}\delta_{\mu}^{\nu}-4\mathcal{L}%
_{\mathcal{F}}F_{\mu\lambda}F^{\nu\lambda}\right)  . \label{H12}%
\end{equation}
In spherically symmetric spacetime, we consider a line-element of the form%
\begin{equation}
ds^{2}=-n\left(  r\right)  dt^{2}+\frac{dr^{2}}{f\left(  r\right)  }+L\left(
r\right)  ^{2}\left(  d\theta^{2}+\sin^{2}\theta d\varphi^{2}\right)  ,
\label{H13}%
\end{equation}
where $n\left(  r\right)  ,$ $f\left(  r\right)  ,$ and $L\left(  r\right)  $
are to be obtained. The reference metric tensor can be chosen as%
\begin{equation}
f_{\gamma\beta}=diag\left[  0,0,h\left(  r\right)  ^{2},h\left(  r\right)
^{2}\sin^{2}\theta\right]  , \label{H14}%
\end{equation}
in which $h\left(  r\right)  ^{2}$ is a coupling function. Having considered
the actual metric \eqref{H4} and the reference metric \eqref{H5}, one finds%
\begin{equation}
\mathcal{K}_{\beta}^{\alpha}=diag\left[  0,0,1-\frac{h\left(  r\right)
}{L\left(  r\right)  },1-\frac{h\left(  r\right)  }{L\left(  r\right)
}\right]  , \label{H15}%
\end{equation}
and consequently%
\begin{equation}
X_{t}^{t}=X_{r}^{r}=-\frac{3L-2h}{L}-\alpha\frac{\left(  3L-h\right)  \left(
L-h\right)  }{L^{2}}-\beta\frac{3\left(  L-h\right)  ^{2}}{L^{2}}, \label{H16}%
\end{equation}
and%
\begin{equation}
X_{\theta}^{\theta}=X_{\phi}^{\phi}=-\frac{3L-h}{L}-\alpha\frac{\left(
3L-2h\right)  }{L}-\beta\frac{3\left(  L-h\right)  ^{2}}{L}. \label{H17}%
\end{equation}
Also, the nonlinear Maxwell equation admits an electric field of the form
\cite{SARA3,65}%
\begin{equation}
E\left(  r\right)  =\frac{2}{b}\left(  1-\frac{1}{\sqrt{1+\frac{qb}{r^{2}}}%
}\right)  , \label{H18}%
\end{equation}
and thus the energy momentum tensor components are explicitly found to be%
\begin{equation}
T_{t}^{t}=T_{r}^{r}=\frac{-E^{2}}{\left(  1-\frac{bE}{2}\right)  ^{2}},
\label{H19}%
\end{equation}
and%
\begin{equation}
T_{\theta}^{\theta}=T_{\phi}^{\phi}=\frac{E^{2}}{1-\frac{bE}{2}}. \label{H20}%
\end{equation}
As $X_{t}^{t}=X_{r}^{r},$ one should impose $G_{t}^{t}=G_{r}^{r}$ which
implies that%
\begin{equation}
\frac{d}{dr}\left(  \frac{n}{f}L^{\prime2}\right)  =0, \label{H21}%
\end{equation}
where a prime denotes the derivative of a function with respect to its
argument. One can easily check that for the case of $n(r)=f(r)$ and $L(r)=r,$
Eq. (\ref{H21}) is satisfied. A substitution into the $tt$ or $rr$ components
of the Einstein's equations yield%

\begin{multline}
\label{IS1}f\left(  r\right)  =1-\frac{2M}{r}+\frac{8r^{2}}{3b^{2}}\left(
1+\frac{qb}{r^{2}}\right)  ^{3/2}-\frac{4q}{b}\left(  1+\frac{2r^{2}}%
{3qb}\right)  +\\
\frac{m_{g}^{2}}{r}\int dr\left(  3\left(  1+\alpha+\beta\right)
r^{2}-2\left(  1+2\alpha+3\beta\right)  hr+h^{2}\left(  \alpha+3\beta\right)
\right),
\end{multline}
in which $M$ is an integration constant. Finally, $\theta\theta$ or $\phi\phi$
components of the Einstein's equation admit a trivial solution for the
function $h\left(  r\right)  =h_{0}$ where $h_{0}$ is a constant parameter.
Setting $L\left(  r\right)  =r$ and using the $rr$ component of the Einstein's
equation, after some algebra, we obtain the following analytical metric
function for the charged dGRT\ black hole in the nonlinear electrodynamics
\begin{multline}
f\left(  r\right)  =1-\frac{2M}{r}+\frac{8r^{2}}{3b^{2}}\left(  1+\frac
{qb}{r^{2}}\right)  ^{3/2}-\frac{4q}{b}\left(  1+\frac{2r^{2}}{3qb}\right)
+\label{H25}\\
m_{g}^{2}\left(  \left(  1+\alpha+\beta\right)  r^{2}-\left(  1+2\alpha
+3\beta\right)  h_{0}r+h_{0}^{2}\left(  \alpha+3\beta\right)  \right)  ,
\end{multline}
in which $M$ is an integration constant. On the other hand, it is also
possible to obtain a second set of solutions by considering $h\left(
r\right)  =\frac{3\beta+2\alpha+1}{\alpha+3\beta}r$. After making
straightforward calculations, one gets the following black hole solution%
\begin{equation}
f\left(  r\right)  =1-\frac{2M}{r}-\left(  m_{g}^{2}\frac{\left(  1+\alpha
^{2}+\alpha-3\beta\right)  }{3\left(  \alpha+3\beta\right)  }+\frac{8}{3b^{2}%
}\right)  r^{2}-\label{H26}\\
\frac{4q}{b}+\frac{8r^{2}}{3b^{2}}\left(  1+\frac{qb}{r^{2}}\right)  ^{3/2}.
\end{equation}

\textcolor{blue}{It is worth noting that the class of solutions obtained in this study is a \textit{special case} of more general solution \cite{myrev1}, which has an
energy momentum tensor of cosmological constant type. The derivation of this generic solution \cite{myrev1} does not lean on the ansatz for the physical and reference metric or the St\"{u}ckelberg field \cite{elberg0,elberg1,elberg2}, apart from their isotropy. Namely, the solution is compatible with arbitrary matter component, including the nonlinear electrodynamics. Therefore, the massive
gravity-induced fluid, which behaves like a cosmological constant, can coexist with the isotropically distributed
matter, which could be an alternative to the \textit{dark energy}. In the literature, there exists other remarkable studies which are based on the generic solution (see for example \citep{myex0,myex1}).}

\section{Charged Scalar Perturbations for Charged dRGT Massive Gravity Black
Holes in Nonlinear Electrodynamics}

In this section, we shall study the thermal radiation of the charged dRGT
massive gravity (coupled with nonlinear electrodynamics) black holes. To this
end, we first consider the massless charged Klein-Gordon equation%
\begin{equation}
\frac{1}{\sqrt{-g}}D_{\mu}\left[  \sqrt{-g}g^{\mu\nu}D_{\nu}\right]  \Psi=0,
\label{S27}%
\end{equation}

where%
\begin{equation}
D_{\mu}=\partial_{\mu}-iqA_{\mu}, \label{S28}%
\end{equation}
in which the electromagnetic potential is defined as%
\begin{equation}
A_{t}=-\frac{2}{b}\left(  r-\sqrt{r^{2}+qb}\right)  ,\text{ }A_{r}=A_{\theta
}=A_{\varphi}=0. \label{S29}%
\end{equation}

Plugging the line-element (\ref{H13}) of the charged dRGT massive gravity
black hole in the Klein-Gordon equation (\ref{S27}), we get
\begin{multline}
\left[  -\frac{1}{f\left(  r\right)  }\partial_{t}^{2}\Psi+\frac{1}{f\left(
r\right)  }q^{2}A_{t}^{2}\Psi+\frac{2iqA_{t}}{f\left(  r\right)  }\partial
_{t}\Psi+\frac{2f}{r}\partial_{r}\Psi+\right. \label{S30}\\
\left.  f%
\acute{}%
\left(  r\right)  \partial_{r}\Psi+f\left(  r\right)  \partial_{r}^{2}%
\Psi+\frac{\cos\theta}{r^{2}\sin\theta}\partial_{\theta}\Psi+\frac{1}{r^{2}%
}\partial_{\theta}^{2}\Psi+\frac{1}{r^{2}\sin^{2}\theta}\partial_{\varphi}%
^{2}\Psi\right]  =0.
\end{multline}

We use the following ansatz for the wave function%
\begin{equation}
\Psi\left(  t,r,\Omega\right)  =e^{i\omega t}\frac{\varphi\left(  r\right)
}{r}Y_{lm}\left(  \Omega\right)  , \label{S31}%
\end{equation}
in which $e^{i\omega t}$ is the oscillating function and $Y_{lm}\left(
\Omega\right)  $ are spherical harmonics, which satisfy the following angular
equation
\begin{equation}
\frac{1}{\sin^{2}\theta}\frac{\partial^{2}Y}{\partial\varphi^{2}}+\frac
{1}{\sin\theta}\left[  \frac{\partial}{\partial\theta}\left(  \sin\theta
\frac{\partial Y}{\partial\theta}\right)  \right]  =-\lambda Y, \label{S32}%
\end{equation}
where $\lambda=l\left(  l+1\right)  $ is the eigenvalue having orbital quantum
number $l$. Thus, the radial equation reads%
\begin{equation}
\frac{f}{\varphi r}\frac{d}{dr}\left[  r^{2}f\frac{d}{dr}\left(  \frac
{\varphi}{r}\right)  \right]  +\left(  \omega-qA_{t}\right)  ^{2}%
-\frac{\lambda f}{r^{2}}=0. \label{S33}%
\end{equation}

The tortoise coordinate is defined by $\frac{dr_{\ast}}{dr}=\frac{1}{f\left(
r\right)  },$which helps us to permute the radial equation to the form of
one-dimensional Schr\"{o}dinger equation%
\begin{equation}
\frac{d^{2}\varphi\left(  r\right)  }{dr_{\ast}^{2}}+\left[  \omega
^{2}-V_{eff}\right]  \varphi\left(  r\right)  =0, \label{S35}%
\end{equation}
where the effective potential in general form for dRGT massive gravity black
holes with nonlinear electrodynamics is defined as%

\begin{equation}
V_{eff}=2\omega qA_{t}-q^{2}A_{t}^{2}+\frac{\lambda f}{r^{2}}+\frac{f}{r}f%
\acute{}%
, \label{SS36}%
\end{equation}
in which $f%
\acute{}%
=\frac{df}{dr}$. Hereafter we split our calculations to the first and second
solution and clarify them by indexes $1$ and $2.$ Let`s rearrange the
Eq. (\ref{H25}) as%

\begin{equation}
f_{1}\left(  r\right)  =1-\frac{2M}{r}+\frac{8r^{2}}{3b^{2}}\left(
1+\frac{qb}{r}\right)  ^{3/2}-\frac{4q}{b}\left(  1+\frac{2r^{2}}{3qb}\right)
+\left(  Ar^{2}-Br+C\right)  , \label{SS37}%
\end{equation}

where%

\begin{align}
A  &  =m_{g}^{2}\left(  1+\alpha+\beta\right),\nonumber\\
B  &  =m_{g}^{2}\left(  1+2\alpha+3\beta\right)  h_{0},\nonumber\\
C  &  =m_{g}^{2}\left(  \alpha+3\beta\right)  h_{0}^{2}.\label{SS38}
\end{align}

By substituting Eq. (\ref{SS37}) and Eq. (\ref{S29}) in the general formula
(\ref{SS36}), then the effective potential for the first solution can be
obtained as
\begin{multline}
V_{eff\left(  1\right)  }=2\omega q(-\frac{2}{b}\left(  r-\sqrt{r^{2}%
+qb}\right)  )-(-\frac{2q}{b}\left(  r-\sqrt{r^{2}+qb}\right)  )^{2}%
+\label{S36}\\
\frac{\lambda}{r^{2}}\left(  1-\frac{2M}{r}+\frac{8r^{2}}{3b^{2}}\left(
1+\frac{qb}{r}\right)  ^{3/2}-\frac{4q}{b}\left(  1+\frac{2r^{2}}{3qb}\right)
+\left(  Ar^{2}-Br+C\right)  \right)  +\\
\frac{1}{r}\left(  1-\frac{2M}{r}+\frac{8r^{2}}{3b^{2}}\left(  1+\frac{qb}%
{r}\right)  ^{3/2}-\frac{4q}{b}\left(  1+\frac{2r^{2}}{3qb}\right)  +\left(
Ar^{2}-Br+C\right)  \right)  \times\\
\left(  \frac{2M}{r^{2}}+\sqrt{1+\frac{qb}{r}}\left(  \frac{16r}{3b^{2}}%
+\frac{4q}{3b}\right)  -\frac{16qr}{3qb^{2}}+2Ar-B\right)  .
\end{multline}

Following the approach of above to derive the effective potential of dRGT
massive gravity with nonlinear electrodynamics for second solution. The metric
function has been introduced by Eq. (\ref{H26}), which we can rewrite it as%

\begin{equation}
f_{2}\left(  r\right)  =1-\frac{2M}{r}-\left(  D+\frac{8}{3b^{2}}\right)
r^{2}-\frac{4q}{b}+\frac{8r^{2}}{3b^{2}}\left(  1+\frac{qb}{r^{2}}\right)
^{3/2}, \label{SK1}%
\end{equation}

where%

\begin{equation}
D=m_{g}^{2}\frac{\left(  1+\alpha^{2}+\alpha-3\beta\right)  }{3\left(
\alpha+3\beta\right)  }.\label{SK2}%
\end{equation}

The effective potential for the second solution is given by%

\begin{multline}
V_{eff\left(  2\right)  }=2\omega q(-\frac{2}{b}\left(  r-\sqrt{r^{2}%
+qb}\right)  )-(-\frac{2q}{b}\left(  r-\sqrt{r^{2}+qb}\right)  )^{2}%
+\label{SK3}\\
\frac{\lambda}{r^{2}}\left(  1-\frac{2M}{r}-\left(  D+\frac{8}{3b^{2}}\right)
r^{2}-\frac{4q}{b}+\frac{8r^{2}}{3b^{2}}\left(  1+\frac{qb}{r^{2}}\right)
^{3/2}\right)  +\\
\frac{1}{r}\left(  1-\frac{2M}{r}-\left(  D+\frac{8}{3b^{2}}\right)
r^{2}-\frac{4q}{b}+\frac{8r^{2}}{3b^{2}}\left(  1+\frac{qb}{r^{2}}\right)
^{3/2}\right)  \times\\
\left(  \frac{2M}{r^{2}}-2r\left(  D+\frac{8}{3b^{2}}\right)  +\sqrt
{1+\frac{qb}{r^{2}}}\left(  \frac{16r}{3b^{2}}-\frac{8q}{3br} \right)  \right).
\end{multline}

\begin{figure}[h]
\centering
\includegraphics[width=9cm,height=10cm]{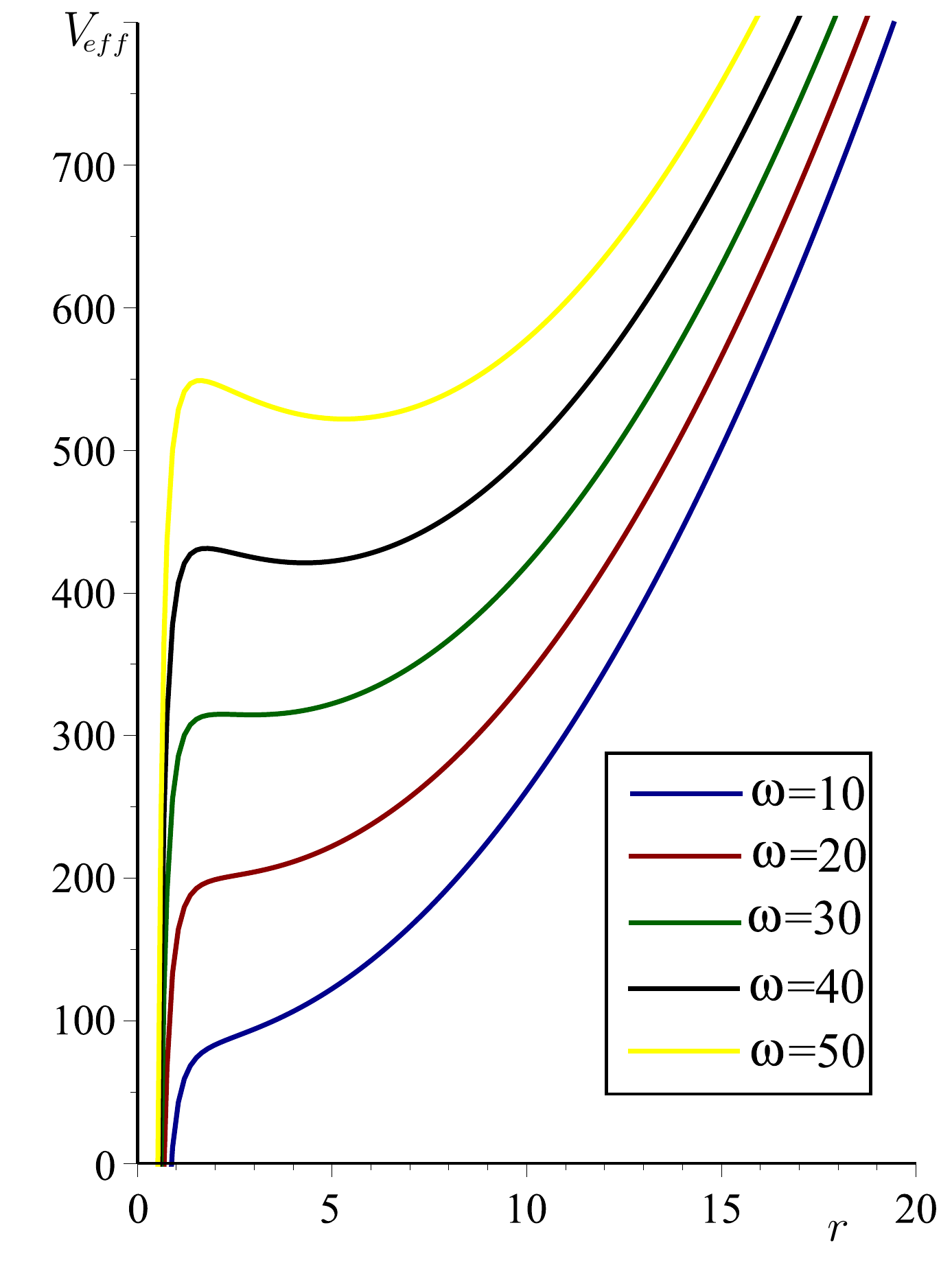}\caption{Plots of $V_{eff}$
versus $r$ for the metric function (\ref{H25}). The plot is governed by Eq.
(\ref{S36}). The physical parameters are chosen as; $M=1, b=50, q=8,$
and $\lambda=0$.}%
\label{myfig1}%
\end{figure}

\begin{figure}[h]
\centering
\includegraphics[width=11cm,height=9cm]{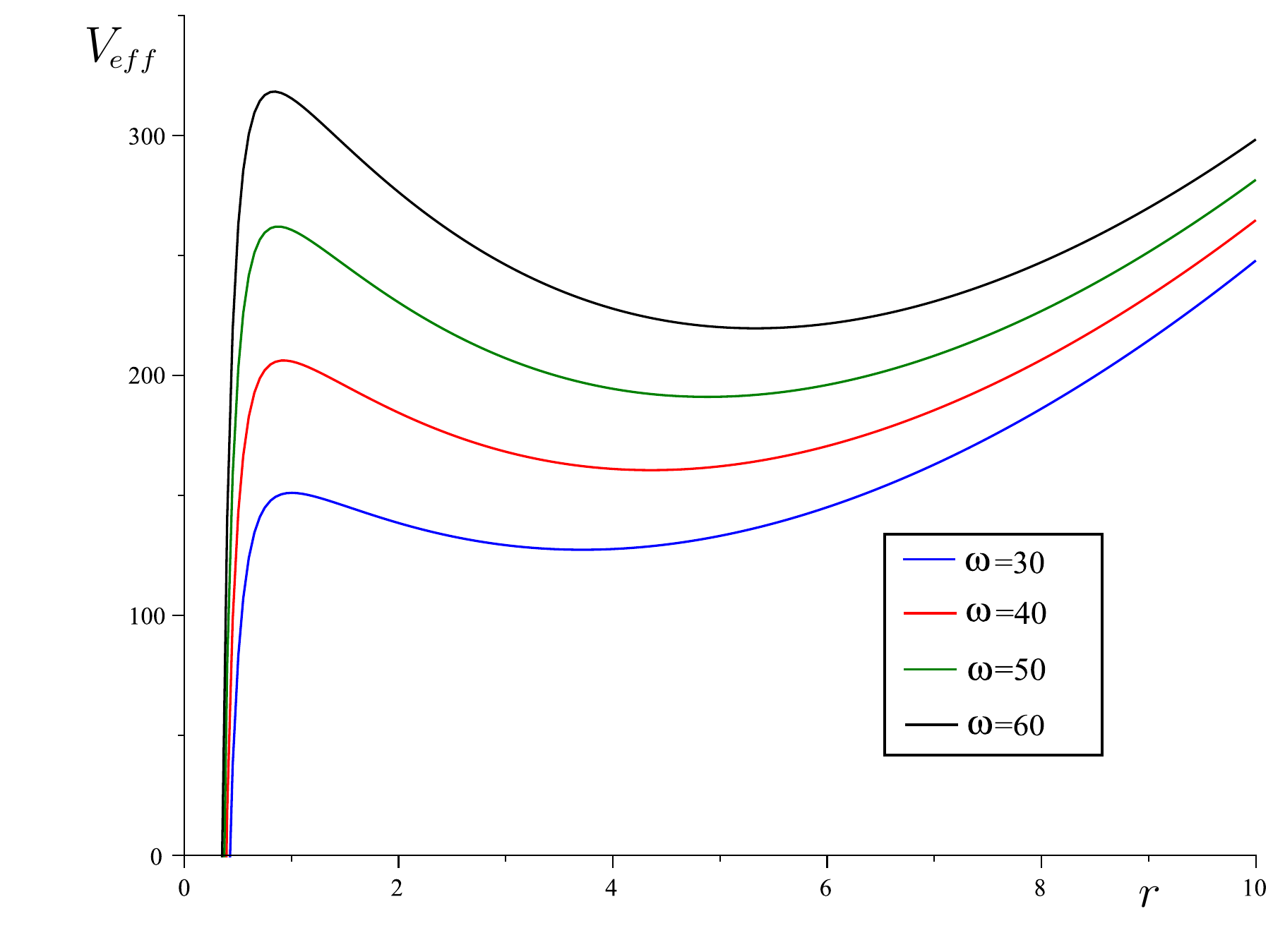}\caption{Plots of $V_{eff}$
versus $r$ for the metric function (\ref{H26}). The plot is governed by Eq.
(\ref{SK3}). The physical parameters are chosen as; $M=1, b=10, q=3,$
and $\lambda=0$.}%
\label{myfig2}%
\end{figure}

The behavior of dRGT effective potential for both solutions i.e., Eqs. (\ref{S36}) and (\ref{SK3})
are depicted in Figs. (\ref{myfig1}) and (\ref{myfig2}) by varying the
controlling parameter of $\omega$ which is appeared in the effective potential
by coupling of nonlinear electrodynamics. The parameters $B$ and $C$ are
chosen to be zero and $A=-1.$ It can be seen from both figures that $V_{eff}%
$, which vanishes at the horizon, peaks right after the horizon and then
quickly dampens towards the asymptotic region, this procedure happened for the
second solution in a smaller amount rather than the first. Moreover for both,
by increasing the frequency the potential peak increase as well. On the other
hand, when the energy of the scalar waves increases, the peak value of the
potential barrier near the event horizon also increases, which may lead to the
caged of the waves. As being stated in Refs. \cite{75,SARA1,77,78}, since the
main contribution to the transmission amplitude comes from the $l=0$ mode
(i.e., $s$-wave case \cite{mychandra}), it is adequate to qualitatively
analyze the potential (\ref{S36}) for $s$-waves. In a general comparison, we
can see the behavior of the potential for second solution is smoother in the
same period than first solution, in this case the role of constant parameter
$b$ is significant.

\section{Rigorous bounds on the greybody factor}

\subsection{First Solution}

In this section, we shall apply the rigorous bounds \cite{74,75} on the
greybody factor to the $3+1$-dimensional black hole in\ dRGT massive gravity
coupled with nonlinear electrodynamics. To this end, we first recall the
formulation of the greybody factor ($T$) \cite{77,78}%
\begin{equation}
T\geq\sec h^{2}\left(  \int_{-\infty}^{+\infty}\vartheta dr_{\ast}\right)  ,
\label{S38}%
\end{equation}
in which
\begin{equation}
\vartheta=\frac{\sqrt{\left[  h%
\acute{}%
\left(  r_{\ast}\right)  \right]  ^{2}+\left[  \omega^{2}-V\left(  r_{\ast
}\right)  -h^{2}\left(  r_{\ast}\right)  \right]  ^{2}}}{2h\left(  r_{\ast
}\right)  }, \label{S39}%
\end{equation}
where $h\left(  r_{\ast}\right)  >0,$ which should satisfy $h\left(
-\infty\right)  =h\left(  +\infty\right)  =\omega.$ Therefore, one can set
$h=\omega$ and hence Eq. (\ref{S39}) simplifies to
\begin{equation}
T\geq\sec h^{2}\left(  \frac{1}{2\omega}\int_{-\infty}^{+\infty}Vdr_{\ast
}\right)  \text{.} \label{S40}%
\end{equation}

By using the tortoise coordinate and the effective potential
of first solution (Eq. (\ref{S36})), then the greybody factor equation
(\ref{S40}) can be written as%

\begin{multline}
T_{1}\geq\sec h^{2}\frac{1}{2\omega}\left\{  \int_{r_{h}}^{R_{h}}\left(
\frac{\lambda}{r^{2}}+\frac{2M}{r^{3}}-\frac{2q^{2}}{r^{4}}+2A-\frac{B}%
{r}\right)  dr+\right. \label{S44}\\
\left.  \int_{r_{h}}^{R_{h}}\frac{2\omega q^{2}r}{Ar^{4}-Br^{3}+\left(
1+c\right)  r^{2}-2Mr+q^{2}}dr-\right. \\
\left.  \int_{r_{h}}^{R_{h}}\frac{\omega q^{3}b}{2\left(  Ar^{5}%
-Br^{4}+\left(  1+c\right)  r^{3}-2Mr^{2}+q^{2}r\right)  }dr\right. \\
\left.  -\int_{r_{h}}^{R_{h}}\frac{q^{4}}{Ar^{4}-Br^{3}+\left(  1+c\right)
r^{2}-2Mr+q^{2}}\right\}  ,
\end{multline}
the result is an awkward formula in point of integration view so to prevail
over this issue we use the Taylor expansion, which accomplish the greybody
factor as
\begin{multline}
T_{1}\geq\sec h^{2}\frac{1}{2\omega}\left\{  -\frac{\lambda}{R_{h}-r_{h}%
}-\frac{M}{R_{h}^{2}-r_{h}^{2}}+\frac{2q^{2}}{3\left(  R_{h}^{3}-r_{h}%
^{3}\right)  }\right. \label{S45}\\
\left.  -(B+\frac{1}{2}\omega qb)\ln\left(  R_{h}-r_{h}\right)  +W_{1}\left(
R_{h}-r_{h}\right)  +X_{1}\left(  R_{h}^{2}-r_{h}^{2}\right)  \right. \\
\left.  +Y_{1}\left(  R_{h}^{3}-r_{h}^{3}\right)  +Z_{1}\left(  R_{h}%
^{4}-r_{h}^{4}\right)  -P_{1}\left(  R^{5}-r^{5}\right)  \right\}  ,
\end{multline}
where%
\begin{equation}
W_{1}=2A-\frac{\omega bM}{q}-q^{2}, \label{S46}%
\end{equation}

\begin{equation}
X_{1}=(\omega+\frac{\omega b}{4q^{2}}\left(  q\left(  1+c\right)
-\frac{4M^{2}}{q}\right)  -M), \label{S47}%
\end{equation}

\begin{equation}
Y_{1}=-\frac{\omega b}{6q^{2}}\left(  qB-\frac{4M\left(  q^{2}\left(
1+c\right)  -2M^{2}\right)  }{q^{3}}\right)  +\frac{q^{2}\left(  1+c\right)
-4M^{2}+4\omega M}{3q^{2}}, \label{SS48}%
\end{equation}
and%

\begin{multline}
Z_{1}=\left.  \frac{\omega}{2q^{2}}\left(  -\left(  1+c\right)  +\frac{4M^{2}%
}{q^{2}}\right)  \right.  -\label{z49}\\
\left.  \frac{1}{8q^{2}}\left(  -\omega qbA+\frac{\omega b\left(  1+c\right)
^{2}}{q}+\frac{4\omega bM\left(  Bq^{4}-3Mq^{2}\left(  1+c\right)
+4M^{3}\right)  }{q^{5}}\right)  \right. \\
\left.  -\frac{1}{4q^{2}}\left(  q^{2}B+\frac{4M\left(  -q^{2}\left(
1+c\right)  +2M^{2}\right)  }{q^{2}}\right)  \right.  ,
\end{multline}
and
\begin{multline}
P_{1}=\left.  \frac{\omega b}{5q^{5}}\left(  \left(  -q^{2}\left(  1+c\right)
+6M^{2}\right)  B-2MAq^{2}\right)  \right.  -\label{S50}\\
\left.  \frac{2\omega bM(1+c)}{5q^{7}}\left(  -q^{2}\left(  1+c\right)
+2M^{2}\right)  \right.  -\\
\left.  \frac{\omega bM}{5q^{9}}\left(  -q^{4}\left(  1+c^{2}\right)
+12M^{2}q^{2}\left(  1+c\right)  -2cq^{4}-16M^{4}\right)  \right.  +\\
\left.  \frac{1}{5q^{2}}\left(  -q^{2}A+4BM+\left(  1+c\right)  ^{2}%
-\frac{4M^{2}\left(  3q^{2}\left(  1+c\right)  -4M^{2}\right)  }{q^{4}%
}\right)  \right.  -\\
\left.  \frac{2}{q^{2}}\left(  \omega B+\frac{4\omega M\left(  -q^{2}\left(
1+c\right)  +2M^{2}\right)  }{q^{4}}\right)  \right.  ,
\end{multline}
two parameters $R_{h}$ and $r_{h}$, are upper and lower rigorous bound
respectively, which they can obtained by \cite{30}%

\begin{equation}
R_{h}=\frac{2}{\left(  -2A\right)  ^{1/3}}\left[  \sqrt{\frac{2\sqrt{3}}%
{\beta}+4}\cos\left(  \frac{1}{3}\sec^{-1}\left(  -\frac{\sqrt{\frac{\sqrt{3}%
}{\beta}+2}\left(  2\sqrt{2}\beta+\sqrt{6}\right)  }{5\beta+3\sqrt{3}}\right)
\right)  -1\right]  , \label{S51}%
\end{equation}
and the lower one reads%
\begin{equation}
r_{h}=\frac{-2}{\left(  -2A\right)  ^{1/3}}\left[  \sqrt{\frac{2\sqrt{3}%
}{\beta}+4}\cos\left(  \frac{1}{3}\sec^{-1}\left(  -\frac{\sqrt{\frac{\sqrt
{3}}{\beta}+2}\left(  2\sqrt{2}\beta+\sqrt{6}\right)  }{5\beta+3\sqrt{3}%
}\right)  +\frac{\pi}{3}\right)  +1\right]  . \label{S52}%
\end{equation}

We demystify our results obtained, by illustrating the greybody factors for
different charge values, in this case to approach in ideal form of figure we
got a $\ $significantly smaller amount of $b\left(  \text{around }0.1\right)
$ than its value $\left(  b=50\right)  $ in effective potential case$.$ The
remarkable point in Fig. (3) is that greybody factor for $h_{0}=0$ behaves as
in the case of the AdS/dS black string (\cite{30,SARA1}). From this figure, one
can see that the greyboday factor increase by only increasing a small range of
charge but after it has an inverse behaviour.

\subsection{Second Solution}

Based on previous part, let us substitute the effective potential of second
solution Eq. (\ref{SK3}) in Eq. (\ref{S40}) to get,%
\begin{multline}
T_{2}\geq\sec h^{2}\frac{1}{2\omega}\left\{  \int_{r_{h}}^{R_{h}}\left(
\frac{\lambda}{r^{2}}+\frac{2M}{r^{3}}-\frac{2q^{2}}{r^{4}}-2D\right)
dr+\right. \\
\left.  \int_{r_{h}}^{R_{h}}\frac{2\omega q^{2}r}{-Dr^{4}+r^{2}-2Mr+q^{2}%
}dr-\int_{r_{h}}^{R_{h}}\frac{\omega q^{3}b}{2\left(  -Dr^{5}+r^{3}%
-2Mr^{2}+q^{2}r\right)  }dr\right. \\
\left.  -\int_{r_{h}}^{R_{h}}\frac{q^{4}}{-Dr^{4}+r^{2}-2Mr+q^{2}}dr\right\},
\end{multline}
then after integration and using the Taylor expansion, the greybody equation
is defined as%
\begin{multline}
T_{2}\geq\sec h^{2}\frac{1}{2\omega}\left\{  -\frac{\lambda}{R_{h}-r_{h}%
}-\frac{M}{R_{h}^{2}-r_{h}^{2}}+\frac{2q^{2}}{3\left(  R_{h}^{3}-r_{h}%
^{3}\right)  }-\right. \label{S56}\\
\left.  \frac{1}{2}\omega qb\ln(R_{h}-r_{h})-W_{2}(R_{h}-r_{h})+X_{2}\left(
R_{h}^{2}-r_{h}^{2}\right)  +\right. \\
\left.  Y_{2}\left(  R_{h}^{3}-r_{h}^{3}\right)  +Z_{2}\left(  R_{h}^{4}%
-r_{h}^{4}\right)  +P_{2}\left(  R_{h}^{5}-r_{h}^{5}\right)  \right\}  ,
\end{multline}
where%

\begin{equation}
W_{2}=2D+q^{2}+\frac{\omega bM}{q}, \label{S57}%
\end{equation}

\begin{equation}
X_{2}=\omega-\left(  \frac{-\omega b}{4q}+\frac{\omega bM^{2}}{q^{3}}\right)
-M, \label{S58}%
\end{equation}

\begin{equation}
Y_{2}=\frac{q^{2}-4M^{2}+4\omega M}{3q^{2}}+\frac{2\omega bM\left(
q^{2}-2M^{2}\right)  }{3q^{5}} \label{S59}%
\end{equation}

\begin{equation}
Z_{2}=\frac{\omega}{2q^{2}}\left(  \frac{4M^{2}}{q^{2}}-1\right)
-\frac{\omega b}{8q^{2}}\left(  -qD+\frac{\left(  q^{2}-12M^{2}\right)
}{q^{3}}+\frac{16M^{4}}{q^{5}}\right)  , \label{S60}%
\end{equation}
and%

\begin{multline}
P_{2}=-\frac{1}{10q^{3}}\left(  2\omega bM\left(  2D+\frac{3}{q^{2}}%
-\frac{16M^{2}}{q^{4}}+\frac{16M^{4}}{q^{6}}\right)  \right)  -\label{S61}\\
\frac{1}{5q^{2}}\left(  1-q^{2}D+\frac{4M}{q^{4}}\left(  M\left(
4M^{2}-3q^{2}\right)  +2\omega\left(  -q^{2}+2M^{2}\right)  \right)  \right)
.
\end{multline}

\begin{figure}[h]
\centering
\includegraphics[width=10cm,height=11cm]{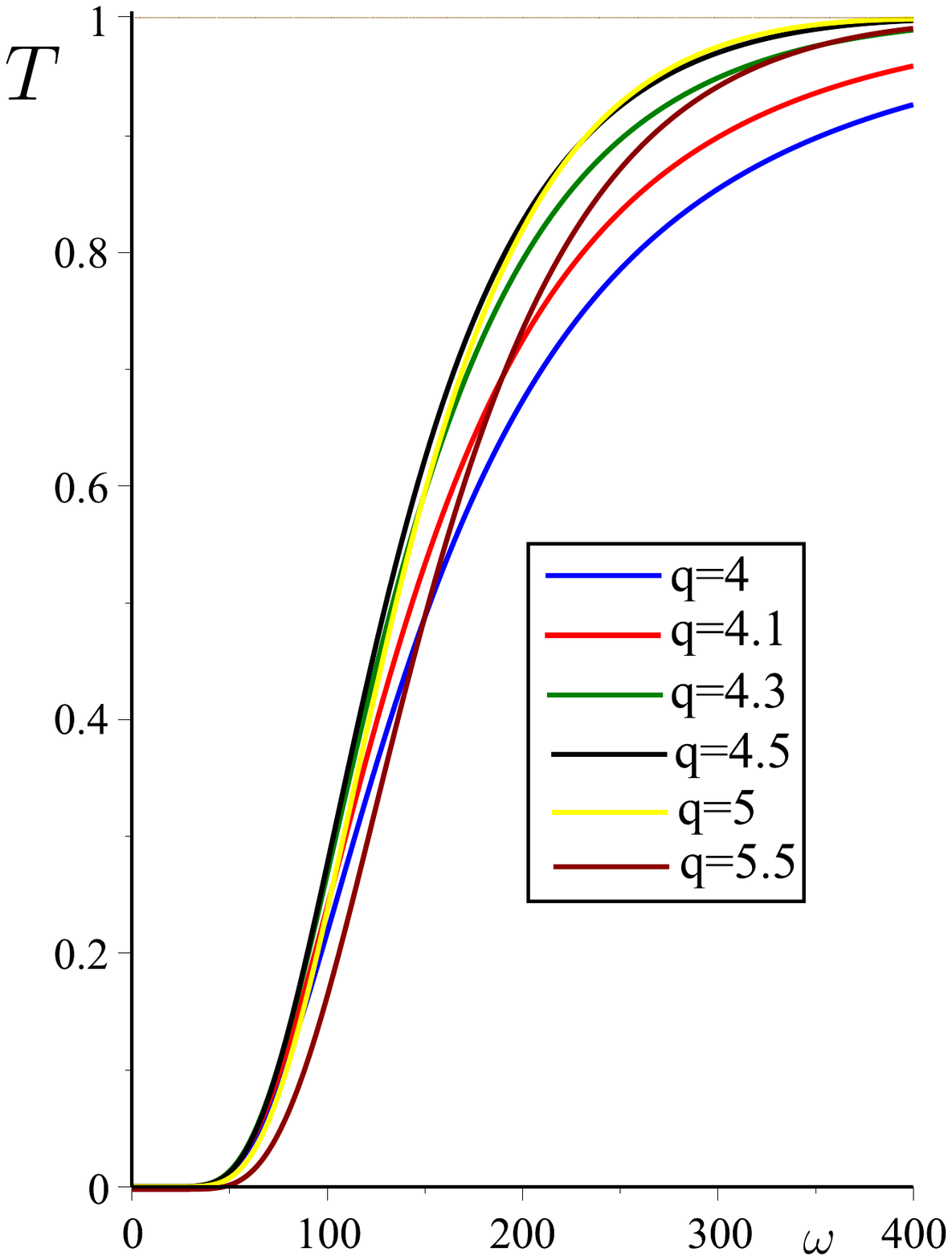}\caption{Plots of $T$ versus
$\omega$ for the metric function $f_{1}$. The plot is governed by Eq.
(\ref{S56}). The physical parameters are chosen as; $M=1, b=0.1, \lambda=0, A=-1,$ and $B=C=0$.}%
\label{myfig3}%
\end{figure}

\begin{figure}[h]
\centering
\includegraphics[width=13cm,height=11cm]{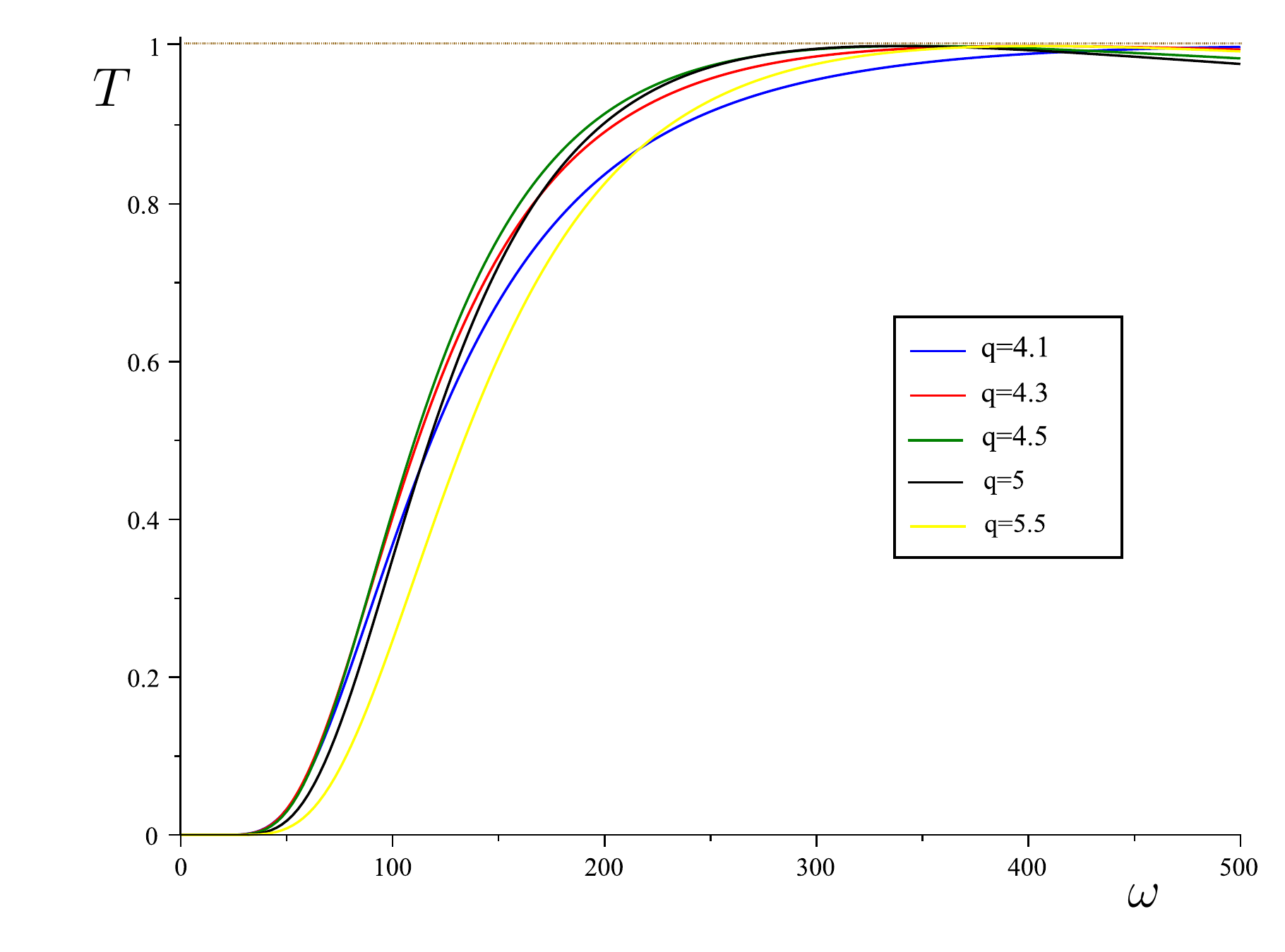}\caption{Plots of $T$
versus $\omega$ for the metric function $f_{2}$. The plot is governed by Eq.
(\ref{S56}). The physical parameters are chosen as; $M=1, b=0.1, \lambda
=0,$ and $D=0.8$.}%
\label{myfig4}%
\end{figure}

From the Eq. (\ref{S56}), one can see the rigorous bounds on the greybody
factors for the second solution of dRGT massive gravity coupled with nonlinear
electrodynamics and its plotted as shown in Fig. (4), the constant parameter
$b$ is chosen to be small in comparison with the potential case, for both
solution of greybody factor. We can see that by increasing the charge
parameter gradually, the greybody factor approach to its maximum value then it
starts to dwindle, this alteration happen after $q=4.5$ for both solutions.
Therefore we can conclude that, having a monotonous behaviour in existence of
charge for greybody factor is far from expectation.

\section{Conclusions}

In this study, we have first sought for the dRGT black holes in nonlinear
electrodynamics. We have shown that there exists two possible class of
$3+1$-dimensional black solutions in\ the dRGT massive gravity coupled with
nonlinear electrodynamics. The obtained spacetimes admit static and
spherically symmetric metric (\ref{H13}). However, each class of the charged
dRGT\ black holes has different metric functions (\ref{IS1}) and (\ref{H25})
depending on the considered coupling functions $h\left(  r\right)  =h_{0}$ and
$h\left(  r\right)  =\frac{3\beta+2\alpha+1}{\alpha+3\beta}r$, which are
emerged from the Einstein's field equations, respectively. We have then
derived the effective potential through the decoupled set of radial and
angular equations resulting from the massless charged Klein-Gordon equation.
The behaviors of the effective potential for both solutions have been depicted
in Figs. (1) and (2) for the metric functions $f_{1}$ and $f_{2}$,
respectively. The impression of this utility method in greybody radiation is
illustrated in the Figs. (3) and (4). In fact, the greybody factors in dRGT
massive gravity with linear electrodynamics was studied in (\cite{30,75,SARA1}).
Thus, we have the revealed the influence of other parameters, which are
consequences of coupling with nonlinear electrodynamics, on the potential and
greybody factor for the two different black holes solutions.

The greybody factors that we are interested in are just the transmission
probabilities for scalar wave modes propagating through the effective
potential. We have managed to obtain several rigorous bounds that are placed
on the greybody factors of the charged dRGT black holes. In particular, we
have seen that the structure of the effective potential is deterministic for
the rigorous bound on the greybody factor. Furthermore, we have depicted the
greybody factors, which are derived from the rigorous bound. Based upon our
analysis, we have seen that charged dRGT black holes of nonlinear
electrodynamics evaporate quickly as compared to the charged dRGT black holes
originated from linear electrodynamics \cite{36}. Namely, nonlinear
electrodynamics gives rise to the dRGT black holes radiate more thermal flux
of quantum particles. For this reason, they will disappear in a shorter time
than the charged one belonging to the linear electrodynamics.

Since the rotating black hole solutions in modified gravity theories are
significant as they offer an arena to test these theories through
astrophysical observations, in the near future, we plan to obtain the rotating
dGRT black holes having charge in nonlinear electrodynamics by using the
standard Newman-Janis algorithm \cite{76} and reveal the effect of the
rotation on their evaporation.

\section*{Acknowledgements}

The authors are grateful to the Editor and anonymous Referees for their
valuable comments and suggestions to improve the paper.

\end{document}